\documentclass[twocolumn,prb, showpacs]{revtex4-1}
\usepackage{graphicx}
\usepackage{dcolumn}
\usepackage{bm}
\usepackage{color}

\begin{document}


\title{Revisiting the crystal electric field of P2-$Na_xCoO_2$\\ for the intermediate spin state of Co$^{3+}$}

\author{G. J. Shu$^1$}
\author{F. C. Chou$^{1,2}$}
\email{fcchou@ntu.edu.tw}
\affiliation{
$^1$Center for Condensed Matter Sciences, National Taiwan University, Taipei 10617, Taiwan}
\affiliation{
$^2$National Synchrotron Radiation Research Center, Hsinchu 30076, Taiwan}

\date{\today}

\begin{abstract}
The magnetic moment per Co$^{4+}$ of P2($\gamma$)-type Na$_x$CoO$_2$ in Curie-Weiss metal regime is revisited and examined under a newly proposed modification of the octahedral crystal electric field (CEF).  The proposed model explains the origin of the existence of the intermediate state S=1 for Co$^{3+}$ through an exciton-like elementary excitation of the narrow gap between the split $e_g$ and $t_{2g}$ groups.  The CoO$_2$ layer is proposed to be constructed from the tilted edge-sharing square-planar CoO$_2$ chains with inter-chain coupling.  The square-planar CEF of CoO$_2$ requires covalent bond formation between Co and the four in-plane neighboring oxygens, while the oxygens sitting in the neighboring chains can be viewed as apical oxygens of low effective charge for a CoO$_6$ pseudo-octahedron.  The reason why angle-resolved photoemission spectroscopy (ARPES) failed to observe the local-density approximation (LDA) predicted $e_g'$ hole pockets at $k_z$=0 and the reversed order of the $t_{2g}$ splittings between LDA and CEF calculations can also be resolved using the proposed model.       

\end{abstract}


                             
\maketitle


\begin{figure}
\includegraphics[width=3.5in]{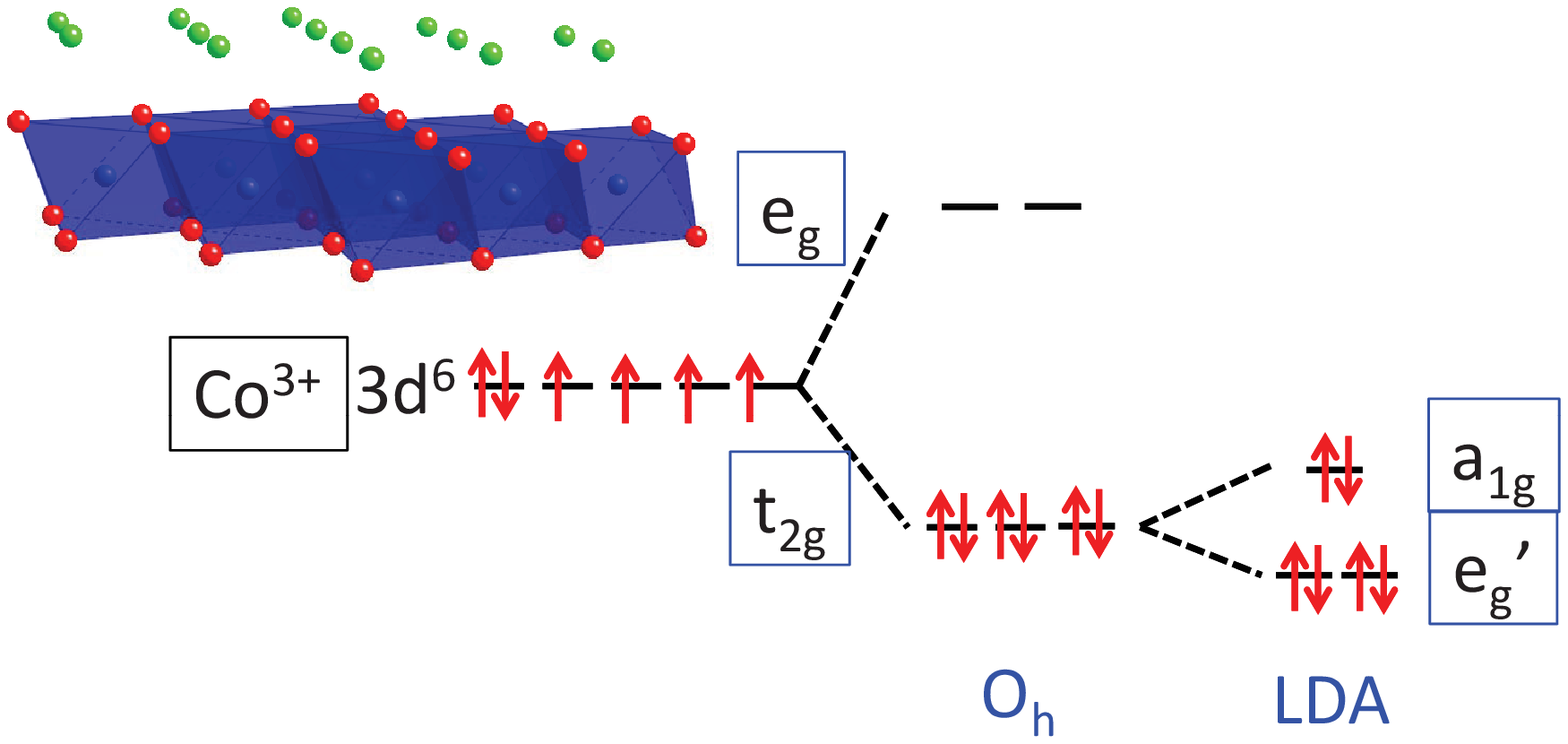}
\caption{\label{fig-NaxCoO2}(color online) The crystal structure of Na$_x$CoO$_2$ is presented with the CoO$_2$ and Na layers in the space group $P6_3/mmc$.  The conventional electronic configuration of Co$^{3+}$ (3d$^6$) in an octahedral crystal field splits into $e_g-t_{2g}$ groups, and the additional splitting of $t_{2g}$ ($e_g'-a_{1g}$) is due to trigonal distortion. In contrary to the spectrum of lower $a_{1g}$ predicted from the CEF calculation, the $a_{1g}$ is higher based on LDA calculations.\cite{Singh2000, Landron2008}}
\vspace{-5mm}
\end{figure}

Layered P2-type (or $\gamma$-type) Na$_x$CoO$_2$ is an important class of material for both fundamental and applied aspects, e.g., it is a good candidate thermoelectric material with a high thermoelectric figure-of-merit (ZT) at high temperature;\cite{Terasaki1997} has a great similarity to the important battery cathode electrode material Li$_x$CoO$_2$;\cite{Hertz2008}, has superconductivity, as observed in Na$_{1/3}$CoO$_2$ after hydration;\cite{Takada2003} and has an intriguing Curie-Weiss metal behavior as a strongly correlated electron material.\cite{Chou2004}  The crystal structure of Na$_x$CoO$_2$ can be viewed as a Na layer sandwiched between CoO$_2$ triangular lattice layers as illustrated in Fig.~\ref{fig-NaxCoO2}. A 2D triangular lattice of CoO$_2$ formed by the closely packed CoO$_6$ octahedra is commonly observed in many cobalt oxide compounds in contrast to the high T$_c$ cuprate materials, which contain a 2D square lattice of CuO$_2$.  The CoO$_2$ layer has been the basic building block of many layered cobalt oxide compounds, including Na$_x$CoO$_2$, Pb$_2$Sr$_2$Co$_2$O$_y$, and Co$_3$Co$_4$O$_9$, of various types of staging and the unavoidable incommensurability is due to layer mismatch.\cite{Sootsman2009}  The physical properties of materials with a hole-doped CoO$_2$ layer have often been interpreted from the mixed valence of Co$^{3+}$ and Co$^{4+}$.  However, the early d-orbital energy levels for Co under the octahedral crystal field assumption (see Fig.~\ref{fig-NaxCoO2}) cannot be used to reasonably interpret the Curie-Weiss metal behavior, especially concerning the puzzling questions of why the theoretically predicted $e_g'$ hole pockets have not been observed experimentally, and why the $t_{2g}$ splitting ($a_{1g}-e_g'$) is reversed between predictions from LDA and CEF calculations.\cite{Zhou2005, Landron2008}    

The existence of an intermediate spin (S=1) for Co$^{3+}$ has been proposed from the ellipsometry experimental results for Na$_{0.82}$CoO$_2$.\cite{Bernhard2004} Daghofer \textit{et al.} proposed that Hund's rule coupling across a narrow gap between the split $e_g$ and $t_{2g}$ groups could permit the existence of S=1 within the proposed spin-orbit-polaron model, i.e., the intermediate spin (IS) state of S=1 for Co$^{3+}$ may exist in addition to the original low spin (LS) state of S=0.\cite{Daghofer2006} However, in view of the large gap in $\Delta$ of $\sim$1 eV between $e_g-t_{2g}$,\cite{Johannes2005} it is difficult to construct one octahedral CEF that has an actual severe distortion. That is, the oxygens must be removed from the original apical positions, which is structurally or chemically nearly impossible due to the CoO$_2$ layer formed by the closely packed edge-sharing octahedra.  In addition, the energy levels of $t_{2g}$ splitting predicted from the \textit{ab initio} calculations are strangely reversed relative to the CEF prediction.\cite{Landron2008}  Landron \textit{et al.} have carefully examined this issue but failed to uncover connections with the metal-ligand hybridization, the long-range crystalline field, the screening effects, and the orbital relaxation; the tentative conclusion still vaguely points to the mixing of $e_g$ and $t_{2g}$.\cite{Landron2008}   

\begin{figure}
\includegraphics[width=3.5in]{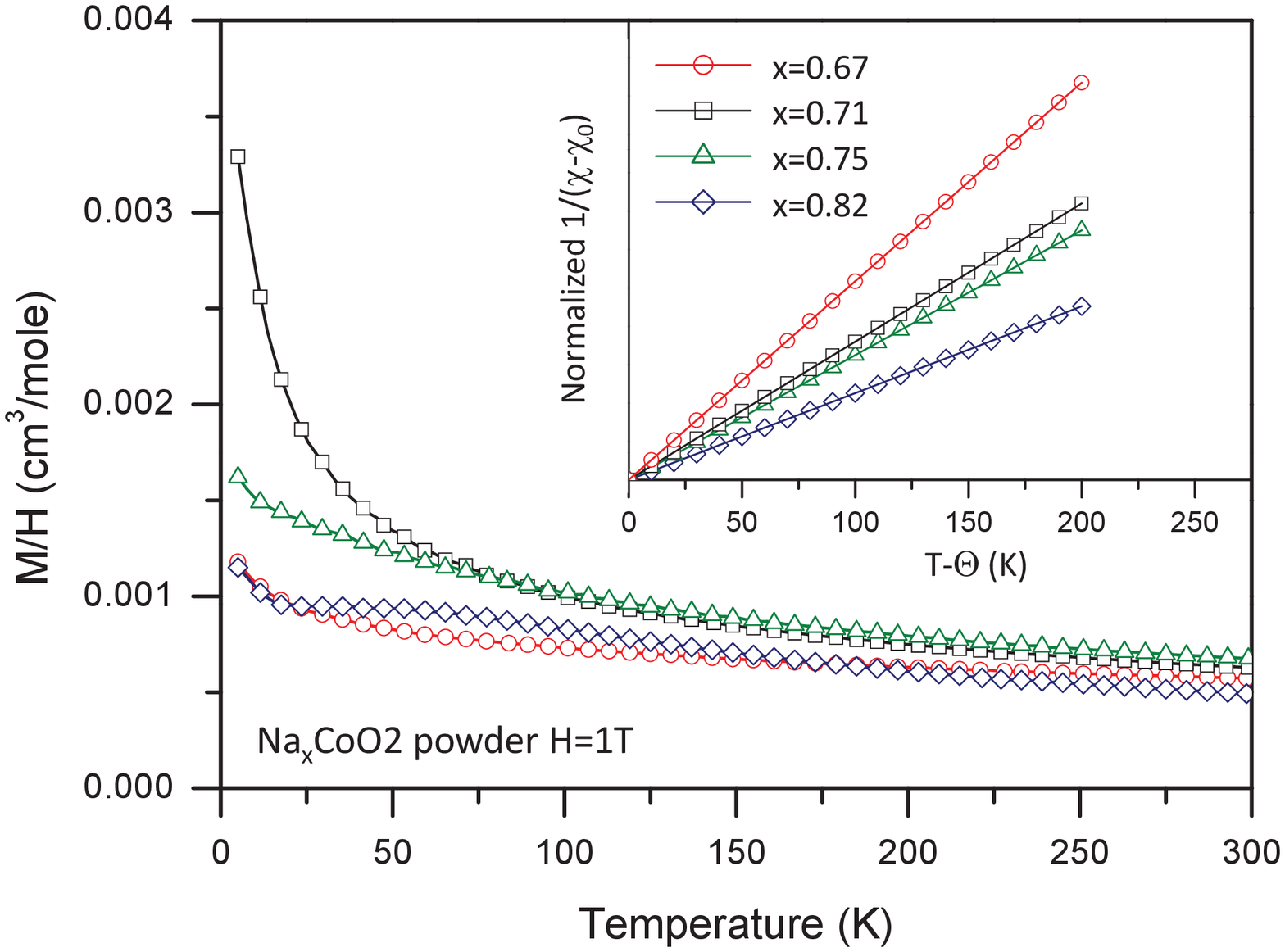}
\caption{\label{fig-chiT}(color online) Average spin susceptibilities of Na$_x$CoO$_2$ (x=0.67-0.82) in a 1 Tesla applied field.  The inset shows normalized $1/(\chi$-$\chi_{\circ}$)  vs. (T-${\Theta}$) to illustrate the x-dependence of the Curie constant by the different slopes corresponding to 1/C. }
\vspace{-5mm}
\end{figure}

\begin{table}
\caption{\label{table-mueff} Curie-Weiss law, $\chi(T)=\chi_\circ + C/(T-\Theta)$, fitting for polycrystalline samples of $Na_xCoO_2$ in an applied field H= 1 Tesla. }
\begin{tabular}{c|c|c|c|c}
 \hline
x  & 0.67& 0.71& 0.75 & 0.82  \\
 \hline
fitting range  & 60-300K  & 65-300K & 60-300K & 125-300K\\
$\chi_\circ$ (cm$^3$/mole)  & 0.0004(5)& 0.0003(7) & 0.0003(6)& 0.0001(5) \\
C (cm$^3$-K/mole)  &  0.0675(5) & 0.1082(1)  & 0.1311(3) & 0.1423(2) \\
$\Theta$(K) & -103.5(7) & -70.2(5) & -106.1(7) & -102.5(3)    \\
$\mu_{eff}$ ($\mu_B$ per Co) &  0.7350 & 0.9305 & 1.0243 & 1.0671   \\
$\mu_{eff}$ ($\mu_B$ per Co$^{4+}$) &  1.279 & 1.728 & 2.049 & 2.515    \\
$\alpha$ & -0.084* & 0.00068 & 0.049 & 0.091\\ 
 \hline
\end{tabular}
*This value is unphysical, which implies that the
itinerant contribution is no longer negligible.
\end{table}

We have revisited this issue experimentally through the analysis of the effective magnetic moment for a series of Na$_x$CoO$_2$ samples in the Curie-Weiss metal regime.  Single crystal samples of Na$_x$CoO$_2$ of x$\sim$0.8 were grown using the optical floating-zone method in an oxygen atmosphere.  The Na contents of the single crystal samples were fine-tuned to x=0.67, 0.71, 0.75, and 0.82 using an electrochemical intercalation technique and verified with electron microprobe analysis (EPMA), within an error of $\pm$0.01, as described previously.\cite{Shu2007}  Average spin susceptibility ($\chi$(T)=M/H) data were acquired from a pulverized powder sample under an applied magnetic field of 1 Tesla in the temperature range of 5-300K, as shown in Fig.~\ref{fig-chiT}.  The spin susceptibilities can be fitted satisfactorily with the Curie-Weiss law, $\chi(T)=\chi_\circ + C/(T-\Theta)$, in the paramagnetic regime and are summarized in Table~\ref{table-mueff}.    Assuming the amount of Co$^{4+}$ ($(t_{2g})^5$) with S=1/2 can be determined directly from the Na content x without oxygen vacancies, it is expected that the $\mu_{eff}$ values per Co$^{4+}$ should be 1.732 $\mu_B$ with g=2.  However, it is clear that except for the sample of x=0.71, the $\mu_{eff}$ per Co$^{4+}$ values are progressively and significantly higher than 1.732 $\mu_B$ for x $>$ 0.71, as indicated in Table~\ref{table-mueff}.  Similar results have also been reported consistently in all early published works,\cite{Chou2004, Wang2003, Luo2004, Rhyee2008} including for Li$_x$CoO$_2$, which has a similar CoO$_2$ 2D triangular lattice.\cite{Hertz2008}  While a Na content higher than $\sim$0.71 is expected to contain a lower level of Co$^{4+}$ and less doped itinerant holes, the increasing Curie constant remains unclear.

\begin{figure*}
\includegraphics[width=4.5in]{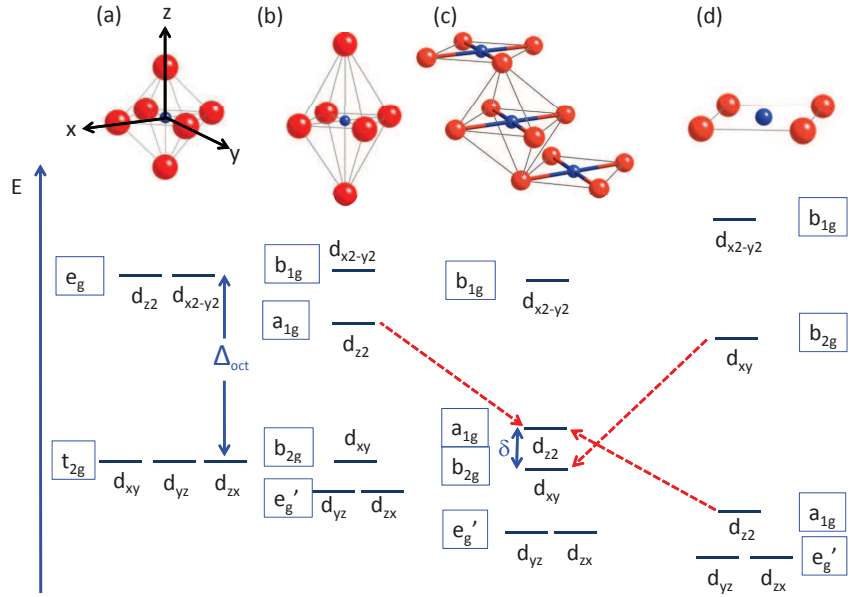}
\caption{\label{fig-CEF}(color online) The d-orbital energy splittings of a transition metal surrounded by oxygen ligands in (a) regular octahedral CEF in O$_h$ symmetry, (b) octahedral CEF with principal z-axis elongation in D$_{4h}$ symmetry, (c) the proposed CEF with weak effective charge at the apical ligand positions, and (d) the square-planar CEF in C$_{4v}$ symmetry.\cite{Moore} }
\vspace{-5mm}
\end{figure*}

While the CoO$_2$ layer of Na$_x$CoO$_2$ has been viewed as closely packed edge-sharing octahedra that forms a 2D triangular lattice, the distortion from the flattened CoO$_2$ layer along the rhombohedral (111) axis requires the distortion of octahedral CEF, whereupon the $t_{2g}$ degeneracy is lifted.  The octahedral distortion has been reflected on the subtle difference of Co-O bond lengths and O-Co-O bond angles in CoO$_6$ through synchrotron X-ray structure refinement before.\cite{Huang2009}   In the crystal field theory for elongated octahedral CEF, as illustrated in Fig.~\ref{fig-CEF}, $e_g$ splits into $b_{1g}$-$a_{1g}$ and $t_{2g}$ splits into $b_{2g}$-$e_g'$, where``a/b", ``e", and ``t" denote single, double, and triple degeneracy in the convention of group theory, respectively.\cite{Moore}  The two groups of the $e_g$ doublet and $t_{2g}$ triplet have a large gap of a $\Delta_{oct}$ of $\sim$1 eV, and the elongation of the apical Co-O distance lifts the degeneracy of both $e_g$ and $t_{2g}$ by lowering the z-related levels within each group.  The extreme case of such a distortion can be observed in the square-planar CEF that is derived by removing apical oxygens to infinity, which lowers the z-related levels significantly, as demonstrated in Fig.~\ref{fig-CEF}(d).  However, since the first published LDA band calculation for Na$_x$CoO$_2$, the $t_{2g}$ splitting has been commonly mislabeled as $a_{1g}$-$e_g'$ in the physics community (see Fig.~\ref{fig-NaxCoO2}),\cite{Singh2000} and $a_{1g}=(d_{xy}+d_{yz}+d_{zx})/\sqrt{3}$ and $e_g'=[(d_{zx}-d_{yz})/\sqrt{2}, (2d_{xy}-d_{yz}-d_{zx})/\sqrt{6}]$ after an axis transformation of the principal z-axis from the apical oxygen direction to the (111) direction in rhombohedral symmetry.\cite{Zhou2005, Landron2008}  Note that the relative positions of the ligand does not change after axis transformation, and an identical $t_{2g}$ splitting, as a result of octahedral CEF distortion, can be described for any selection of an axes system.  The most common distortion in the rhombohedral symmetry description is the thickness change of the CoO$_2$ layer, which can be roughly represented by a small elongation from the original perfect CoO$_6$ octahedron.      

\begin{figure}
\includegraphics[width=3.5in]{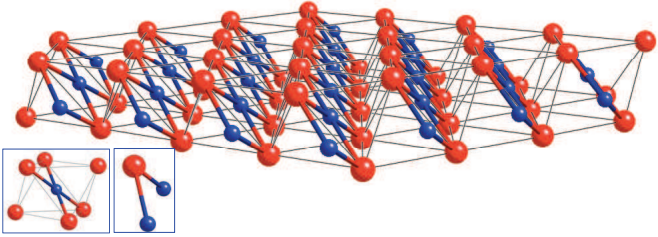}
\caption{\label{fig-chain}(color online) The CoO$_2$ plane can also be viewed as edge-sharing square-planar chains with inter-chain coupling instead of edge-sharing octahedra. The pseudo-octahedron of CoO$_6$ and Co-O-Co at 90$^\circ$ are shown in the insets.}
\vspace{-5mm}
\end{figure}

If the CoO$_2$ layer is viewed as closely packed edge-sharing octahedra, it is impossible to create one distorted octahedral CEF by adjusting the apical Co-O bond lengths alone, i.e., to create a distortion severe enough for both the $e_g$ and $t_{2g}$ groups to split wide and leave one small gap between them.  Instead of assuming that the CoO$_2$ layer which is composed of edge-sharing octahedra, we propose an alternative view; the CoO$_2$ layer could be composed of edge-sharing square-planar CoO$_2$ chains with nontrivial inter-chain coupling, as illustrated in Fig.~\ref{fig-chain}.  There are two main reasons to make such an assumption; the first is because the earlier view of octahedral CEF with trigonal distortion failed to provide a reasonable argument to allow the formation of a narrow gap between the $e_g$ and $t_{2g}$ groups, and the second is that each oxygen with two electrons missing from its $2p$-orbital can form, at most, two covalent bonds with the neighboring Co atoms in a 90$^\circ$ coordination, as illustrated in the inset of Fig.~\ref{fig-chain}.  Chemically, it is impossible for oxygen to form three covalent bonds with the neighboring three Co atoms, as implied in the earlier octahedral picture.

The modified view of CEF is effectively in between an elongated octahedral CEF and a square-planar CEF.  It is possible that the Co-O only has four covalent $\sigma$-bonds within each square-planar CoO$_2$ unit, and the apical oxygens should not be viewed as individual ligands but as weak effective charges dressed by the neighboring edge-sharing chains, as illustrated in Fig.~\ref{fig-chain}.  The CoO$_2$ plane originally assumed to be in edge-sharing octahedra is now viewed as edge-sharing square-planar chains with inter-chain coupling, i.e., all of the oxygens in each square-planar edge-sharing chain also serve as the effective apical oxygen for the neighboring square-planar CoO$_2$ unit to form a pseudo-octahedron.  The actual crystal field can now be understood as a unique octahedral CEF with apical oxygens of a much less effective charge.  In fact similar energy splitting can also be generated through the addtional magnetic couplings of $J_{diag}$ and $J'$ per polaron unit defined in the calculations by Daghofer \textit{et al.}\cite{Daghofer2006}  Starting from the square-planar CEF, the effective charge of a distorted octahedral CEF may be formed when the $a_{1g}$ (d$_{z^2}$) level is raised slightly above the $b_{2g}$ (d$_{xy}$) level.  A clear level inversion between $a_{1g}$ and $b_{2g}$ between the square-planar and elongated octahedral CEF occurs once the effective charge of the apical oxygen is properly tuned through the inter-chain coupling .

\begin{figure}
\includegraphics[width=3.5in]{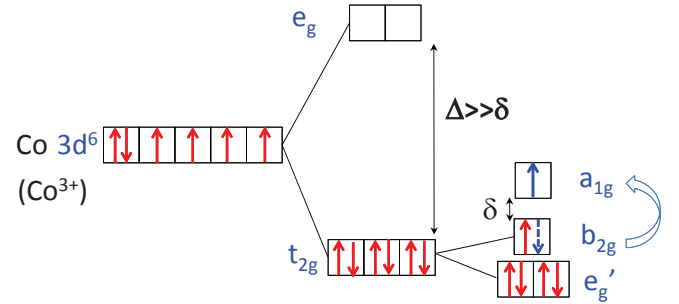}
\caption{\label{fig-spin}(color online) The low spin (LS) state and intermediate spin (IS) state for Co$^{3+}$ (3d$^6$) derived from the proposed CEF.  Thermal or strain energy at ambient temperature would allow activation between $a_{1g}$-$b_{2g}$ of the narrow gap $\delta$ on the order of $\sim$10 meV.}
\vspace{-5mm}
\end{figure}

Following our proposed model, which leads to the possible existence of a narrow gap between the $e_g$ and $t_{2g}$ groups, i.e., the small gap ($\delta$) between $a_{1g}$-$b_{2g}$ shown in Fig.~\ref{fig-CEF}(c), we may re-examine the spins of the 3d electrons in the Co ions.  It is commonly accepted that the LS states of Co$^{3+}$ ($3d^6$) and Co$^{4+}$ ($3d^5$) are S=0  and S=1/2, respectively.  In the  newly constructed CEF with a narrow gap between $a_{1g}$-$b_{2g}$  of $\sim$10 meV,\cite{Zhou2005} the thermal or stain energy at ambient temperature would be sufficient to activate electrons from the filled $b_{2g}$ to the empty $a_{1g}$, as illustrated in Fig.~\ref{fig-spin}.  In fact, Hund's rule coupling is an alternative description for the existence of IS state when the pairing energy is able to overcome the small CEF gap.\cite{Daghofer2006}  For the IS state (S=1) of Co$^{3+}$ generated through the activation process, the quantity and positions of S=1 must fluctuate in a Boltzmann distribution of population and at random positions.  It is reasonable to assume that the Curie constant should change as a function of temperature, i.e., increased S=1 should be observed at higher temperatures because of increased S=1 activation across the narrow $a_{1g}-b_{2g}$ gap for Co$^{3+}$.  

\begin{figure}
\includegraphics[width=3.5in]{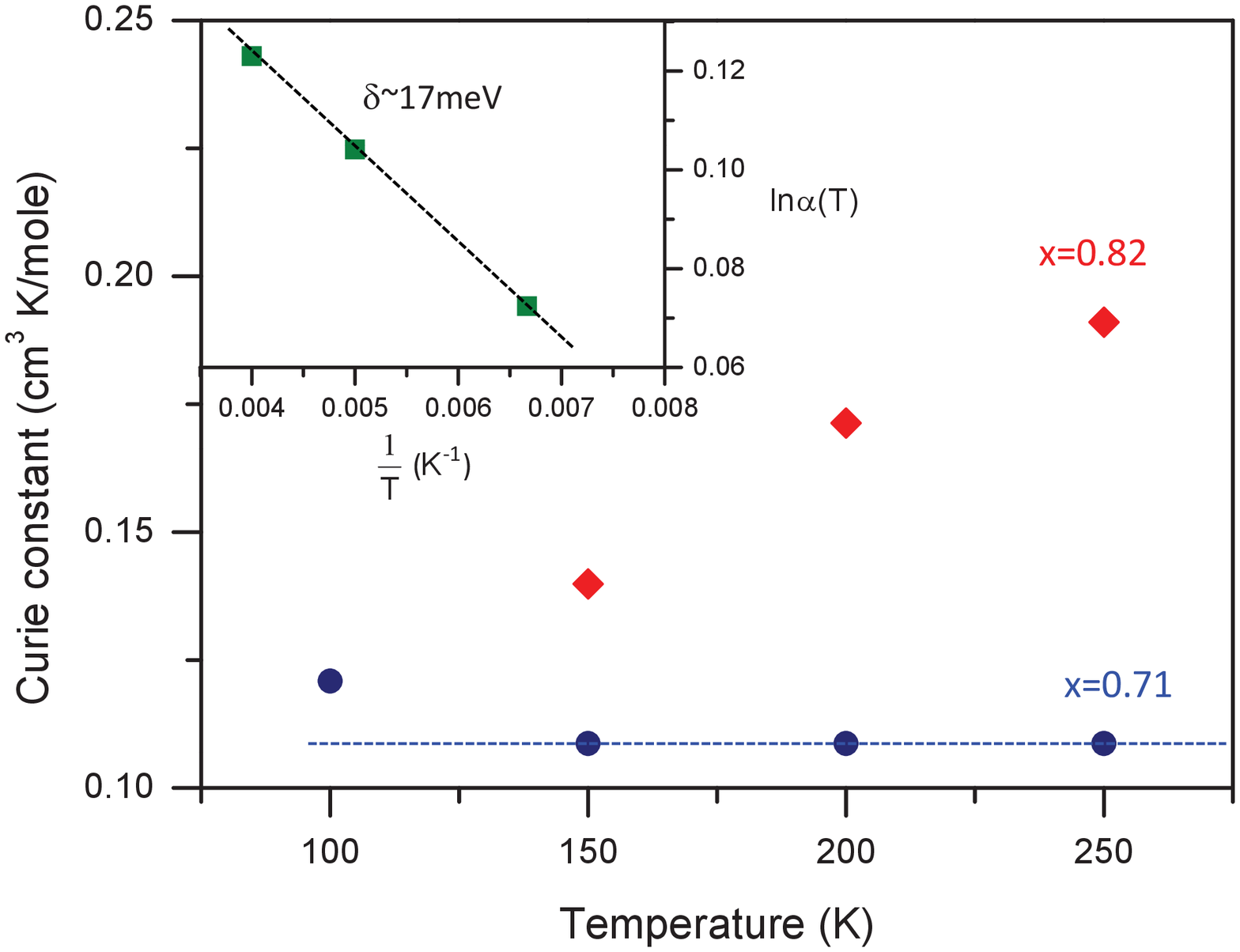}
\caption{\label{fig-CT}(color online) Temperature dependence of Curie constants for x=0.82 and 0.71, where Curie-Weiss law fittings were performed using the selected temperature sections of $\pm$50K centered between 100-250K. The inset shows Arrhenius plot for the number of activated S=1 state of Co$^{3+}$ ($\alpha$(T)) with activation energy ($\delta$) between $a_{1g}-b_{2g}$ fitted to be $\sim$17 meV. }
\vspace{-5mm}
\end{figure}
The Curie-Weiss law for the localized spins is an approximation of Brillouin function under the condition of $\mu H/k_B T \ll1$, i.e., the Curie constant should be temperature independent at low field and high temperature regime.  As a double check to the validity of the applied Curie-Weiss law fitting, the fitted Curie constant for x=0.71 does show a temperature independent value at high temperature above $\sim$150K (see Fig.~\ref{fig-CT}), and only deviates at low temperature when the $\mu H/k_B T\ll1$ condition is violated.  On the other hand, we find that the Curie constant has a subtle temperature dependence when the Curie-Weiss law is fitted using different temperature sections, especially for x=0.82 as shown in Fig.~\ref{fig-CT}. The temperature dependence of C(T) (C=$N\mu_{eff}^2/3k_B$) could be coming from the $N(T)$ through added number of localized spins, or from the $\mu_{eff}(T)$ through added couplings that enhance the size of the magnetic moment of the localized spins effectively.  T. Moriya has shown that the spin fluctuation of FM coupled itinerant electrons in the strongly correlated system can also be described with a Curie's law defined in parallel to that of the localized spins.\cite{MoriyaBook}  While we cannot rule out that the observed Curie constant has contribution from the correlated itinerant electrons completely, especially from the RKKY type coupling of itinerant electrons to the localized spins of S=1/2 for Co$^{4+}$,\cite{Balicas2008} the impact of these couplings should be diminishing at higher temperature.  However, the increasing C(T) as a function of temperature for x=0.82 (see Fig.~\ref{fig-CT}) does not support, or at least implies that the scenario of coupling enhanced effective moment cannot be the dominant factor.  Quite the contrary, the increasing C(T) suggests that more localized spins are added to the system at higher temperature, which agrees with a picture that more spins of S=1 are generated from the thermally activated IS state of Co$^{3+}$.

The confusing excess magnetic moment per Co$^{4+}$ beyond S=1/2 presented in Table~\ref{table-mueff} can now be analyzed quantitatively using the proposed model.  Using $Na_{0.82}CoO_2$ as an example,  the Curie constant (C=N$\mu_{eff}^2$/3$k_B$) has three possible contributions from the LS state of Co$^{4+}$ with S=1/2 ($\mu_{Co^{4+}}^{LS}$=1.732$\mu_B$), the LS state of Co$^{3+}$ with S=0 ($\mu_{Co^{3+}}^{LS}$=0), and the IS state of Co$^{3+}$ with S=1($\mu_{Co^{3+}}^{IS}$=2.828$\mu_B$).  Under the constraint of total N=N$_{Co^{4+}}^{LS}$+N$_{Co^{3+}}^{LS}$+N$_{Co^{3+}}^{IS}$, the $\mu_{eff}$ per Co ion can be analyzed with a modified Curie-Weiss law ($\chi = \chi_\circ + \frac{C}{T-\Theta}$) of
\begin{eqnarray*}
C= \frac{N_{Co^{4+}}^{LS}(\mu_{Co^{4+}}^{LS})^2+N_{Co^{3+}}^{LS}(\mu_{Co^{3+}}^{LS})^2+N_{Co^{3+}}^{IS}(\mu_{Co^{3+}}^{IS})^2}{3k_B}.
\label{eq:one}
\end{eqnarray*}

\noindent We may estimate the fraction ($\alpha$) of Co$^{3+}$ at the activated IS state (S=1) at any instance from
\begin{eqnarray*}
\mu_{eff}^2 = (1-0.82)\times1.732^2+0.82[(1-\alpha)\times0^2+\alpha\times2.828^2].
\label{eq:one}
\end{eqnarray*}

\noindent The fractions of Co$^{3+}$ at S=1, $\alpha$, is estimated to be nearly 9$\%$ under 1 Tesla in the temperature range of 125-300K (see Table~\ref{table-mueff}), which quantitatively agrees with a description that at most one Co$^{3+}$ in a Na di-vacancy formed $\sqrt{13}$a superlattice (1-$\frac{11}{13}$$\sim$9$\%$) is activated, i.e., for each di-vacancy formed supercell with 2 Co$^{4+}$ at the corners and 11 Co$^{3+}$ in the middle, only one of the Co$^{3+}$ (random in time and position) is activated from S=0 to S=1.  In the meantime, there is almost no IS state activation for x=0.71, as indicated by the $\alpha$$\approx$0 value observed in Table~\ref{table-mueff}, which is reasonable because x=0.71 has a smaller $\sqrt{12}$a superlattice size formed by mixed Na tri- and quadri-vacancies of larger Co$^{4+}$ clusters.\cite{Chou2004, Shu2009}  We must note that the regular magnetic field strength does not provide enough Zeeman energy to keep the electron stabilized at the IS state of S=1; however, the narrow gap activation process would keep a fixed amount of Co$^{3+}$ ions at the IS state at random positions.  In fact, such a narrow gap activation process is equivalent to an elementary excitation, similar to that of the quasiparticle exciton used as a method of energy transport in condensed matter without actual net charge transport.

The temperature dependence of Curie constant C(T) for x=0.82 (Fig.~\ref{fig-CT}) implies that the number of localized spins is raised at higher temperature, which is in accord with the proposed model of thermally activated IS state (S=1) for Co$^{3+}$.  Following the same quantitative analysis that extracts the fraction of Co$^{3+}$ at S=1, $\alpha$, the $a_{1g}-b_{2g}$ gap (i.e., $\delta$ defined in Fig.~\ref{fig-spin}) can be estimated from the fitting of activation energy of $\alpha$(T)$\sim$e$^{(-\delta/k_B T)}$ in Arrhenius law, as shown in the inset of Fig.~\ref{fig-CT} with log scale.  The activation energy for Co$^{3+}$ activated from the ground state of S=0 to the excited state of S=1 between the narrow gap of $a_{1g}-b_{2g}$ is fitted to be $\delta$$\sim$17 meV, which provides a reasonable order experimentally and is in agreement with that estimated from the crystal field calculation.\cite{Zhou2005, Daghofer2006}

The current model can also resolve the puzzling contradiction concerning why LDA calculations predicted $e_g'$ hole pockets along $\Gamma - K$ at $k_z$=0 have never been observed in ARPES experiments.\cite{Singh2000, Yang2007}  A strong electronic correlation has been proposed to be responsible for the missing $e_g'$ hole pockets, i.e., the $e_g'$ band could be pushed below the Fermi level.\cite{Zhou2005}  Based on our proposed CEF model, the absence of $e_g'$ hole pockets is not surprising at all because the $e_g'$ band should be filled in the valence band further below both $a_{1g}$ and $b_{2g}$, which is consistent with the description of a large inter-orbital repulsion U$^\prime$ proposed by Zhou \textit{et al.}\cite{Zhou2005}  The inconsistent $a_{1g}$-$e_g'$ level splitting predicted from \textit{ab initio} calculations and the octahedral CEF with trigonal distortion originated from the mislabeling of $a_{1g}$ as shown in Fig.~\ref{fig-CEF}.  The actual CEF should be viewed as unexpectedly weak ligands in the apical positions of a pseudo-octahedral CEF, and the best starting point of a CEF description is the square-planar CEF followed by an increasing inter-chain coupling between the edge-sharing square-planar chains.  The actual narrow gap is between the $a_{1g}$ (which belongs to the $e_g$ group) and $b_{2g}$ (which belongs to the $t_{2g}$ group) levels, as illustrated in Fig.~\ref{fig-CEF}(c), following the standard crystal field labeling by symmetry.

\section*{Acknowledgment}
We are grateful to Patrick A. Lee for many helpful discussions.  FCC acknowledges support from NSC-Taiwan under project number NSC 101-2119-M-002-007.  GJS acknowledges support from NSC-Taiwan under project number NSC 100-2112-M-002-001-MY3.

\end{document}